\documentclass[fleqn,usenatbib]{mnras}

\usepackage[T1]{fontenc}
\usepackage{ae,aecompl}

\usepackage{graphicx}	% Including figure files
\usepackage{amsmath}	% Advanced maths commands
\usepackage{amssymb}	% Extra maths symbols
\usepackage{subfig}
\usepackage{newtxtext,newtxmath}
\newcommand{\tcool}{\tau_\mathrm{cool}}
\newcommand{\review}[1]{#1}
\defcitealias{Manger+2020}{Paper I}

\title[VSI parameter study II]{High Resolution Parameter Study of the Vertical Shear Instability II: Dependence on temperature gradient and cooling time}

\author[N. Manger et al.]{
Manger, Natascha $^{1}$\thanks{E-mail: nmanger@flatironinstitute.org},
Pfeil,Thomas$^{2,3}$ and
Klahr, Hubert$^{2}$
\\
$^{1}$Center for Computational Astrophysics, Flatiron Institute, 162 Fifth Ave, New York, NY 10010, USA\\
$^{2}$Max Planck Institute for Astronomy, K{\"o}nigstuhl 17, 69117 Heidelberg, Germany\\
$^{3}$University Observatory, Faculty of Physics, Ludwig-Maximilians-Universit{\"a}t M{\"u}nchen, Scheinerstr. 1, 81679 Munich, Germany
}

\date{Accepted XXX. Received YYY; in original form ZZZ}

\pubyear{2015}

\begin{document}
\label{firstpage}
\pagerange{\pageref{firstpage}--\pageref{lastpage}}
\maketitle

\begin{abstract}
A certain appeal to the alpha model for turbulence and related viscosity in accretion disks was
that one scales the Reynolds stresses simply on the thermal pressure, assuming that turbulence driven by a certain mechanism will attain a characteristic Mach number in its velocity fluctuations.
Besides the notion that there are different mechanism driving turbulence and angular momentum transport in a disk, we also find that within a single instability mechanism, here the Vertical Shear Instability, stresses do not linearly scale with thermal pressure. 
Here we demonstrate in numerical simulations the effect of the gas temperature gradient and the thermal relaxation time on the average stresses generated in the non-linear stage of the instability. We find that the stresses scale with the square of the exponent of the radial temperature profile \review{at least for a range of $d \log T /d \log R = [-0.5, -1]$, beyond which the pressure scale height varies too much over the simulation domain, to provide clear results. Stresses are also} dependent on thermal relaxation times, provided they are longer than $10^{-3}$ orbital periods. 
The strong dependence of viscous transport of angular momentum on the local conditions in the disk (especially temperature, temperature gradient, and surface density/optical depth) challenges the ideas of viscosity leading to smooth density distributions, opening a route for structure (ring) formation and time variable mass accretion.
\end{abstract}


%%%%%%%%%%%%%%%%%%%%%%%%%%%%%%%%%%%%%%%%%%%%%%%%%%

%%%%%%%%%%%%%%%%% BODY OF PAPER %%%%%%%%%%%%%%%%%%

\section{Introduction}
The mechanism by which protoplanetary disks transport angular momentum is still strongly debated, as is the related question of how turbulence is generated in these disks. Advancing our understanding of either informs the other, as observations tell us that disks accrete at a rate that needs disk winds or turbulent processes to sustain it  \citep[e.g.][]{Hartmann+Bae2018}.
Early studies suggested the Magneto-Rotational Instability \citep[MRI,][]{Balbus+Hawley1991} as a likely driver for turbulence in protoplanetary disks. The MRI however only develops if the assumptions of ideal magnetohydrodynamics are satisfied, chiefly the assumption of an ideal coupling between the gas and the magnetic field. But the coupling between the gas and the magnetic field has subsequently been shown to be weak in large parts of the disk, especially within the regions important for understanding planet formation \citep[see e.g.][]{Turner+2014}. 

This prompted a search for purely hydrodynamic instabilities that can present a viable alternative route to turbulence  \citep{Lyra+Umurhan2019}. Recent works have proposed several possible contenders, among them the Goldreich-Schubert-Fricke instability \citep[GSF,][]{Goldreich+Schubert1967,Fricke1968,Urpin+Brandenburg1998,Arlt+Urpin2004}, later renamed Vertical Shear Instability (VSI) in the context of rotating disks \citep{Nelson+2013}, the convective overstability \citep{Klahr+Hubbard2014,Lyra2014} and its non-linear cousin, the  subcritical baroclinic instability \citep[SBI,][]{Klahr+Bodenheimer2003,Lesur+Papaloizou2010} and the 
zombie vortex instability \citep[ZVI,][]{Marcus+2015,Marcus+2016}.

Of these instabilities, the VSI has recently received the most attention. The instability grows in disk regions where the thermal relaxation timescale is short compared to other timescales and the disk gas couples only weakly to the magnetic field \citep{Nelson+2013,Lin+Youdin2015, Pfeil+Klahr2018}. Hydrodynamical simulations have shown the VSI to generate a turbulent viscosity on the level of $ \alpha \propto 10^{-6} - 10^{-3} $ \citep{Flock+2017,Flock+2020,Stoll+Kley2014,Richard+2016,Manger+Klahr2018,Manger+2020}.
Additionally, the VSI has been shown to support formation of vortices via the RWI \citep{Richard+2016,Manger+Klahr2018,Flock+2020} and the KHI \citep{Latter+Papaloizou2018}, which can concentrate particles locally in the disk and thus form sites where planetesimals can form \citep{Barge+Sommeria1995}. Recent work has also shown the VSI to emerge in non-ideal magnetohydrodynamic simulations with an active magnetically driven wind at the disk surface and a VSI turbulent midplane \citep{Cui+Bai2020}.

However, a clear consensus on the strength of the turbulent viscosity driving angular momentum transport in VSI disks has always been elusive. Models using a simple thermal relaxation prescription showed $ \alpha$ values as diverse as $10^{-6} $ to $10^{-3}$,  while radiation hydrodynamics models showed values on the order of a few $10^{-5} $ to $10^{-4}$. 
But these apparent discrepancies may be explained by considering that the strength of the turbulent viscosity is tied to the underlying disk parameters such as the global temperature and density gradients and may vary with distance to the central star.
\review{Disks that are heated by internal viscous processes and stellar irradiation and which are optically thick can display a wide range of radial slopes of temperature and the disk aspect ratios $H/R$ for constant alpha and mass accretion rates \citep{Belletal1997,Pfeil+Klahr2018}.}
Subsequently, in \citet{Manger+2020}, we found that the total Reynolds stress in the disk scales with $(H/R)^2$ and proposed a theoretical scaling of alpha with both aspect ratio $H/R$ and temperature gradient $q$. Additionally, in \citet{Pfeil+Klahr2020} we showed that the total stresses in the disk varies with the assumed value for the temperature slope of the disk.

In this work, we expand the parameter study started with \citet[hereafter Paper I]{Manger+2020} to systematically investigate the influence of the temperature gradient $q$ and the cooling time $\tau_\mathrm{cool}$ on the Reynolds accretion stresses generated in the disk. We will show that our prediction from \citetalias{Manger+2020} is valid and will further help to reconcile the large range of $\alpha$ values reported form previous VSI simulations conducted under differing disk conditions.
We structure our work as follows. In section 2 we briefly review the disk model and the numerical model used in this work. In section 3 we present the results of our parameter study, which are discussed in section 4. Section 5 presents our conclusions and proposed directions for future work.

\section{Model}
We use a similar disk model and physical setup as in \citetalias{Manger+2020}. All simulations are run using the PLUTO code with a HLLC Riemann Solver with 3rd order piecewise-parabolic reconstruction and a 3rd-order Runge-Kutta time integrator. 

We set the initial conditions in force equilibrium, described by the density profile:
\begin{equation}
\rho = \rho_0 \left(\frac{R}{R_0}\right)^p \exp\left(-\frac{Z^2}{2\,H^2}\right),
\end{equation}
where $R$ and $Z$ represent the radial and vertical coordinate, $H$ is the pressure scale height and $\rho_0$ is the midplane density at reference radius $R_0$. In hydrostatic equilibrium the azimuthal velocity is given by: 
\begin{equation}
v_\phi = \Omega_\mathrm{K}\,R \left[1+q-\frac{q\,R}{\sqrt{R^2+Z^2}}+(p+q)\left(\frac{H}{R}\right)^2\right]^{\frac{1}{2}} \qquad .
\end{equation}
We use a caloric equation of state $\rho e = \frac{P}{\gamma -1}$ and an adiabatic index of $\gamma=5/3$ with the pressure defined as $P = c_\mathrm{s}^2 \rho$ where 
\begin{equation}
    c_\mathrm{s}^2 = c_0^2 \left(\frac{R}{R_0}\right)^q 
\end{equation} 
is the radially varying isothermal sound speed with slope q. In this study, we vary q in the range of $0.4-1.2$, but we set $q$ to the fiducial value $-1.0$ unless otherwise stated. The radial density slope $p=-1.5$ and the disk aspect ratio $H/R=0.1$ are the same across all models, unless specifically noted. Additionally, all models are run with an artificial kinematic viscosity of $ \nu = 10^{-7} $. The complete list of models can be found in table \ref{tab:modelParams}.

To investigate the dependence of the VSI on the cooling time $\tcool$ of the disk gas, we use a simple relaxation scheme for the gas pressure to relax the temperature of the disk to its initial value (described by the isothermal sound speed $c_\mathrm{s,init}$). \review{As the conservative quantity representing thermal energy in the PLUTO code is thermal pressure, we have achieve thermal relaxation by damping the pressure towards the local equilibrium pressure, given by the local density and the desired speed of sound:}
\begin{equation}
\frac{dP}{dt} = -\frac{P-\rho \,c^2_\mathrm{s,init}}{\tau_\mathrm{cool}} \qquad .
\end{equation}
The values of $\tcool$ used are in the range of $10^{-4}$ to $1.0$, with $10^{-4}$ chosen as the fiducial value.

\begin{table}
    \centering
    \begin{tabular}{l l l l}
         Name & $\tcool$ &$q$ & VSI growth \\\hline
         fiducial& $1\cdot 10^{-4}$  & -1.0 & Yes\\\hline % mark as fiducial run
         tau2e-4 & $5\cdot 10^{-4}$ & -1.0 & Yes\\
         tau5e-4 & $5\cdot 10^{-4}$ & -1.0 & Yes\\
         tau1e-3 & $1\cdot 10^{-3}$  & -1.0 & Yes\\
         tau2e-3 & $3\cdot 10^{-3}$ & -1.0 & Yes\\
         tau5e-3 & $5\cdot 10^{-3}$ & -1.0  & Yes\\
         tau1e-2 & $1\cdot 10^{-2}$  & -1.0 & Yes\\
         tau2e-2 & $2\cdot 10^{-2}$ & -1.0 & No\\
         tau5e-2 & $5\cdot 10^{-2}$ & -1.0 & No\\
         tau1e-1 & $1\cdot 10^{-1}$   & -1.0 & No \\
         tau1e0  & $1\cdot 10^{0}$    & -1.0 & No \\\hline
         q-0.4 & $1\cdot 10^{-4}$ & -0.4 & Yes\\
         q-0.5 & $1\cdot 10^{-4}$ & -0.5 & Yes\\
         q-0.6 & $1\cdot 10^{-4}$ & -0.6 & Yes\\
         q-0.7 & $1\cdot 10^{-4}$ & -0.7 & Yes\\
         q-0.8 & $1\cdot 10^{-4}$ & -0.8 & Yes\\
         q-0.9 & $1\cdot 10^{-4}$ & -0.9 & Yes\\
         q-1.1 & $1\cdot 10^{-4}$ & -1.1 & Yes\\
         q-1.2 & $1\cdot 10^{-4}$ & -1.2 & Yes\\\hline
    \end{tabular}
    \caption{List of  models run in this work.}
    \label{tab:modelParams}
\end{table}

All simulations use a common grid with $ n_r \times n_{\theta} \times n_{\phi} = 512 \times 256 \times 64$ grid cells. We use a radial range $0.5-2.0 R_0$ and have a meridional extent of $\pm 3.5 H_0$. Because we are only interested in the average behaviour of the non-linear saturated state of the VSI, we choose a reduced azimuthal extent compared to \review{\citetalias{Manger+2020}} and only use $\phi \in [0-\pi/4]$. Although we showed in \citet{Manger+Klahr2018} that this leads to overall slightly higher $\alpha$ values, we are confident that we can recover the general trend with changing parameters while being able to investigate a larger parameter space.
We employ reflective boundary conditions in the radial and meridional direction and periodic boundaries in the azimuthal direction.

\section{Results}

\subsection{Dependence on cooling time $\tau_\mathrm{cool}$}

To determine in which disk conditions the VSI grows we calculate the $r-\phi$ component of the local Reynolds stress tensor:
\begin{equation}
T_{r,\phi}(r, \theta) \equiv \langle \rho \delta v_r \delta v_\phi \rangle =
\langle \rho v_r v_\phi \rangle - \langle \rho v_r \rangle \langle v_\phi \rangle \,, \label{eqn:ViscStressTens}
\end{equation}
where $\langle ~ \rangle$ denotes an average in azimuth \review{, which by including $\rho$ is automatically mass averaged}. To present $T_{r,\phi}$ in a non-dimenional fashion, we normalise it
by the azimuthally averaged pressure to obtain as intermediate step $\alpha(r, \theta)$,
\begin{equation}
\alpha(r, \theta) = \frac{ T_{r,\phi}}{\langle P \rangle} \qquad.
\end{equation}
which we further mass average to a single paramater for the entire simulation domain as defined by \citet{Shakura+Sunyaev1973}:
\begin{equation}
\label{eq:alpha}
\alpha = \frac{\int \alpha(r, \theta) \rho dV}{\int \rho  dV}\qquad.
\end{equation}

\begin{figure}
    \centering
    \includegraphics[width=\columnwidth]{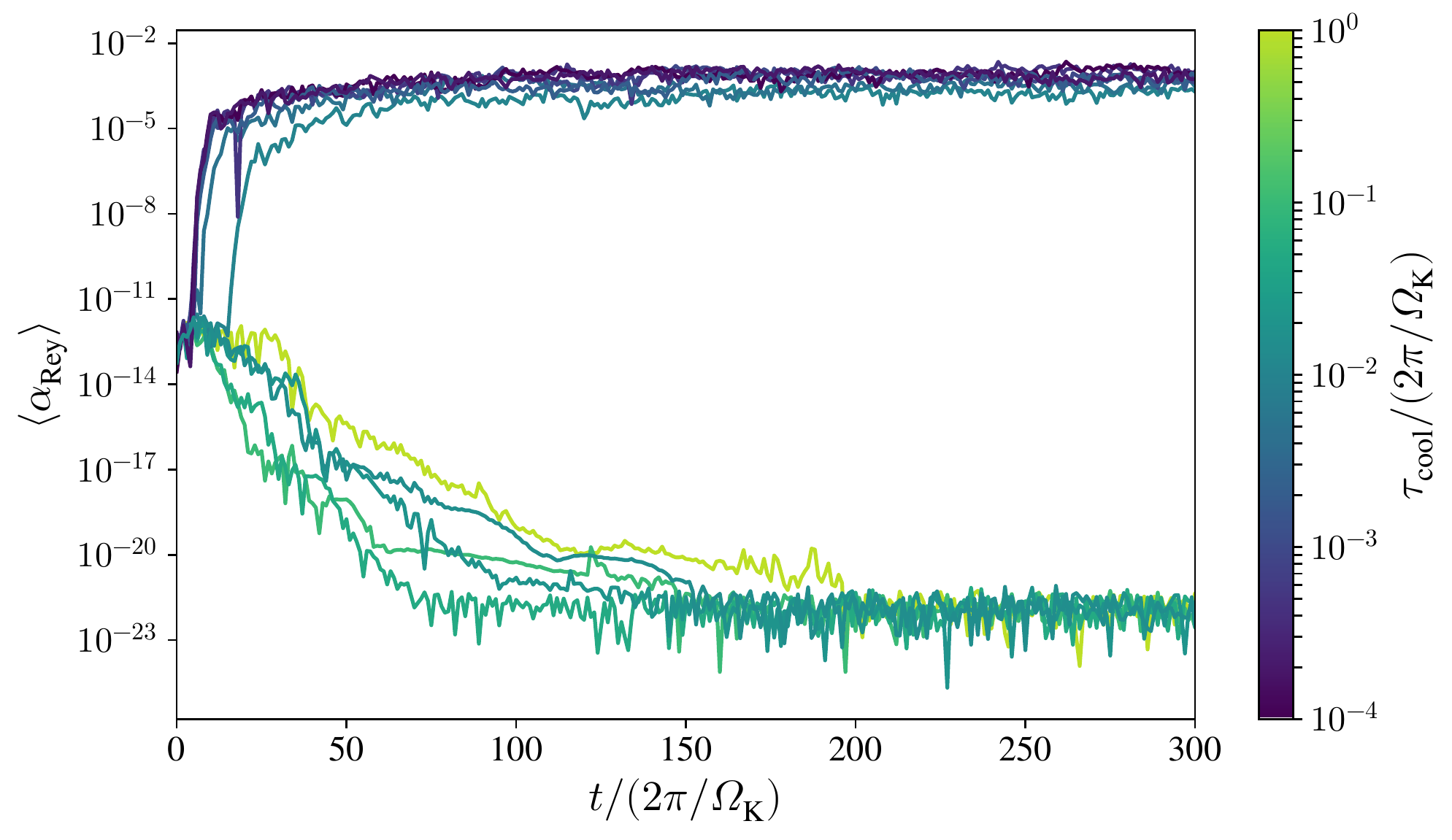}
    \caption{Spatially averaged $\alpha$ values as a function of time. The color represents the cooling time of the individual simulation.}
    \label{fig:alphaTserTau}
\end{figure}

\begin{figure}
    \centering
    \subfloat{\includegraphics[width=\columnwidth]{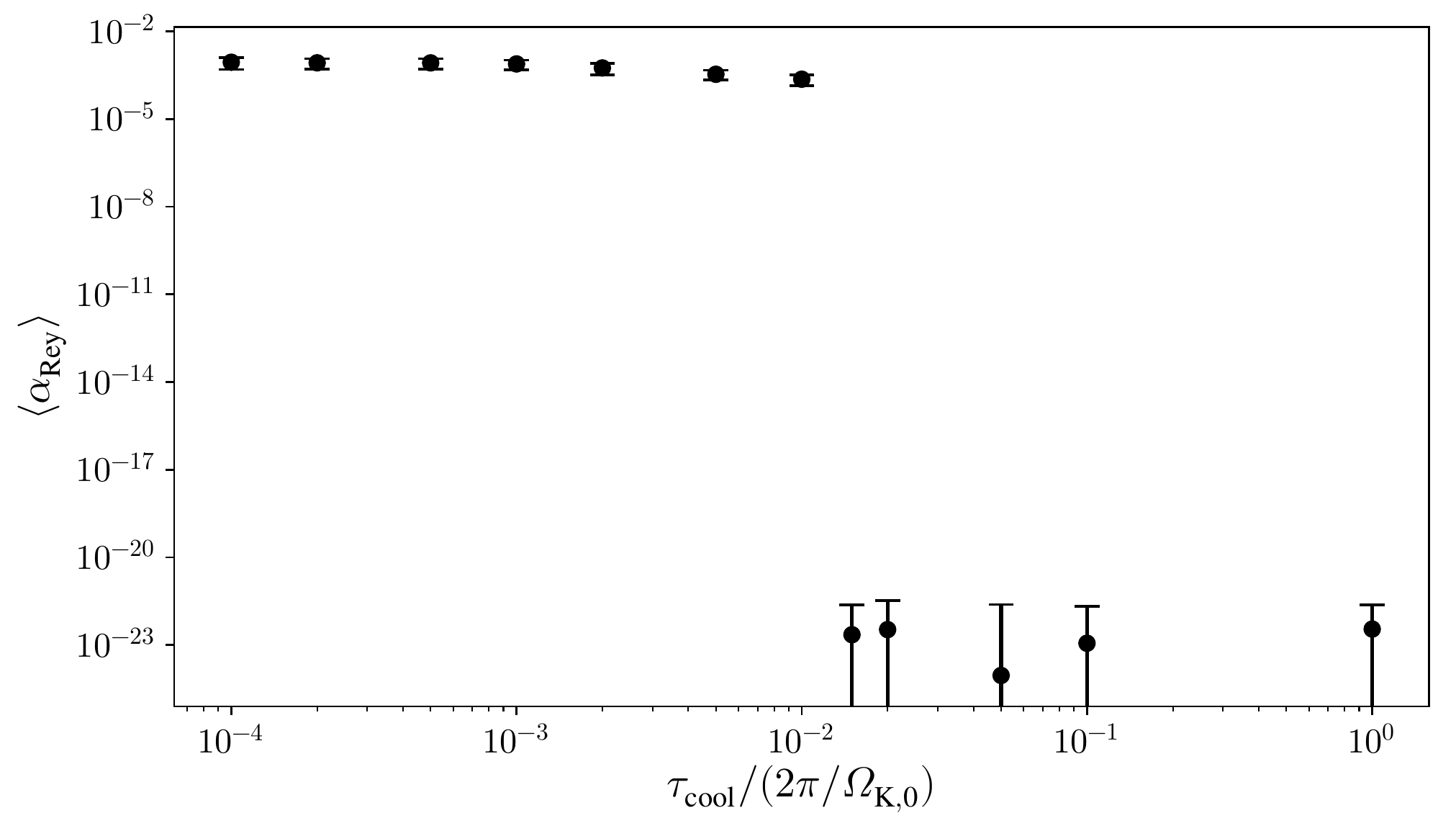}}\\
    \subfloat{\includegraphics[width=\columnwidth]{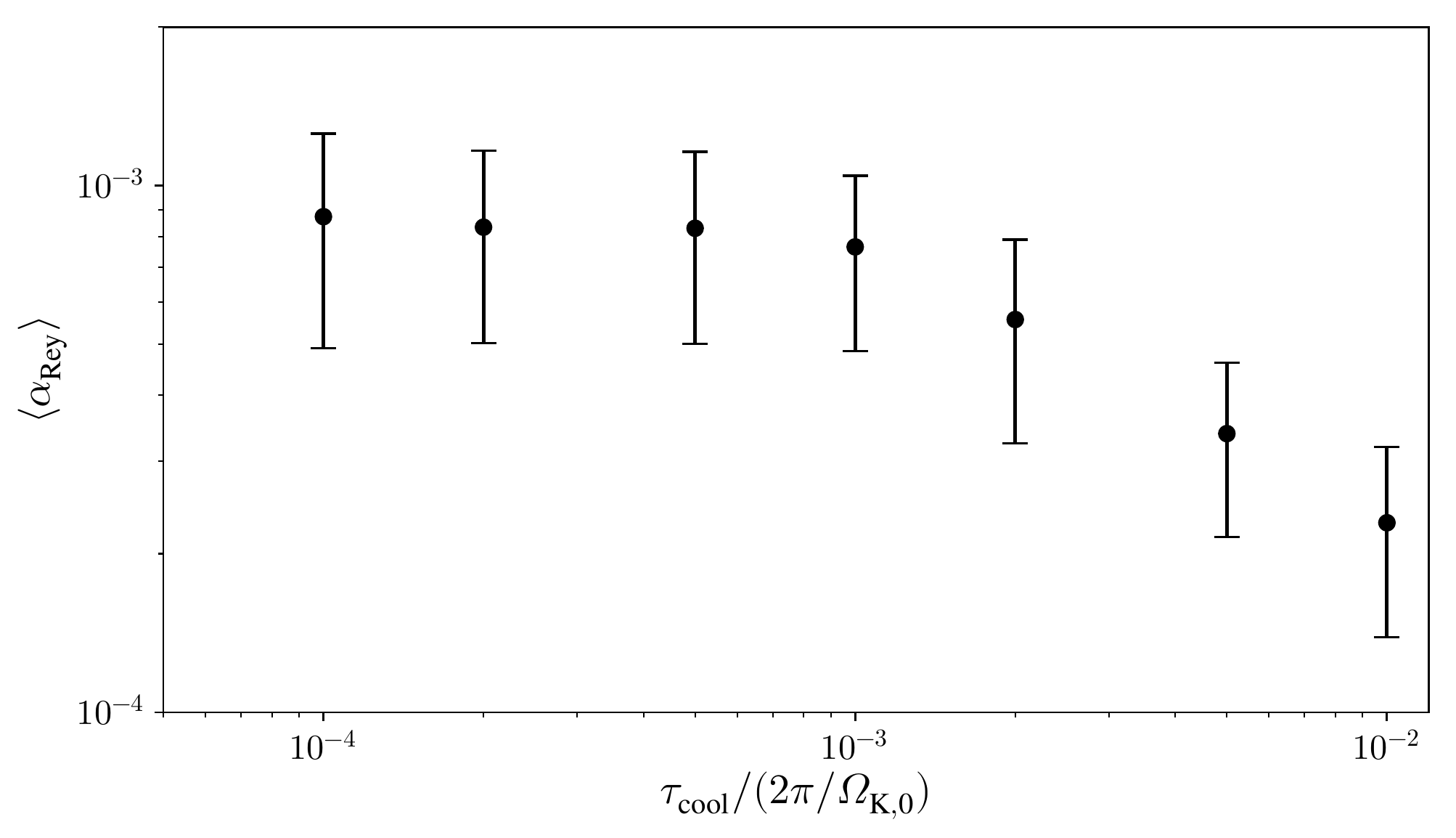}}
    \caption{We plot the space-time average of $\alpha_\mathrm{Rey}$ as function of $\tcool$ for all investigated values (top) and the subset of values enabling VSI growth (bottom). The error bars represent the standard deviation over time only. Clearly visible is a sharp cutoff at $\tcool > 10^{-2}$ as well as a slower decline in $\alpha$ for $10^{-3} < \tcool < 10^{-2}$ orbits.}
    \label{fig:alphaTau}
\end{figure}

We plot the radially and vertically averaged $\alpha$ as a function of time for simulations with different values of $\tcool$ in figure \ref{fig:alphaTserTau}. We find that only part of the simulations show VSI growth, as listed in table \ref{tab:modelParams}.
The top graph of figure \ref{fig:alphaTau} shows the total $\alpha$ value of each simulation from figure \ref{fig:alphaTserTau}, averaged in space and from 200 to 500 reference orbits.

We find a sharp drop in $\alpha$ for simulations at $\tau_\mathrm{cool} \approx 10^{-2}$ orbits, and simulations with $\tcool$ longer than this value do not support VSI growth. This lines up approximately with the critical cooling time of the VSI, given by  \review{\citet{Lin+Youdin2015} as}
\begin{equation}
    \tau_\mathrm{cool,crit}\,\Omega = \frac{\dfrac{H}{R}|q|}{\gamma - 1},
\end{equation}
evaluating to $\tau_\mathrm{cool,crit} = 0.15 \frac{1}{\Omega} = 0.024$ orbits in our case. The fact that the simulation with $\tcool=2\cdot 10^{-2}$ orbits does not show growth is likely due to the viscosity added to the simulation to offset numerical diffusion. Employing a significantly higher resolution would likely enable growth of the VSI, but this would demand a significant amount of computing time and is beyond the scope of this work.

The bottom figure of panel \ref{fig:alphaTau} shows a zoom in on the upper left region of the top figure, showing only the simulations where the VSI grows. In this region of parameter space we find that $\alpha$ values for $\tau_\mathrm{cool} < 10^{-3}$ orbits stay constant, but \review{fall off for longer cooling times, until they reach $\tau_\mathrm{cool,crit}$ and turbulence vanishes completely}.

\begin{figure}
    \centering
    \includegraphics[width=\columnwidth]{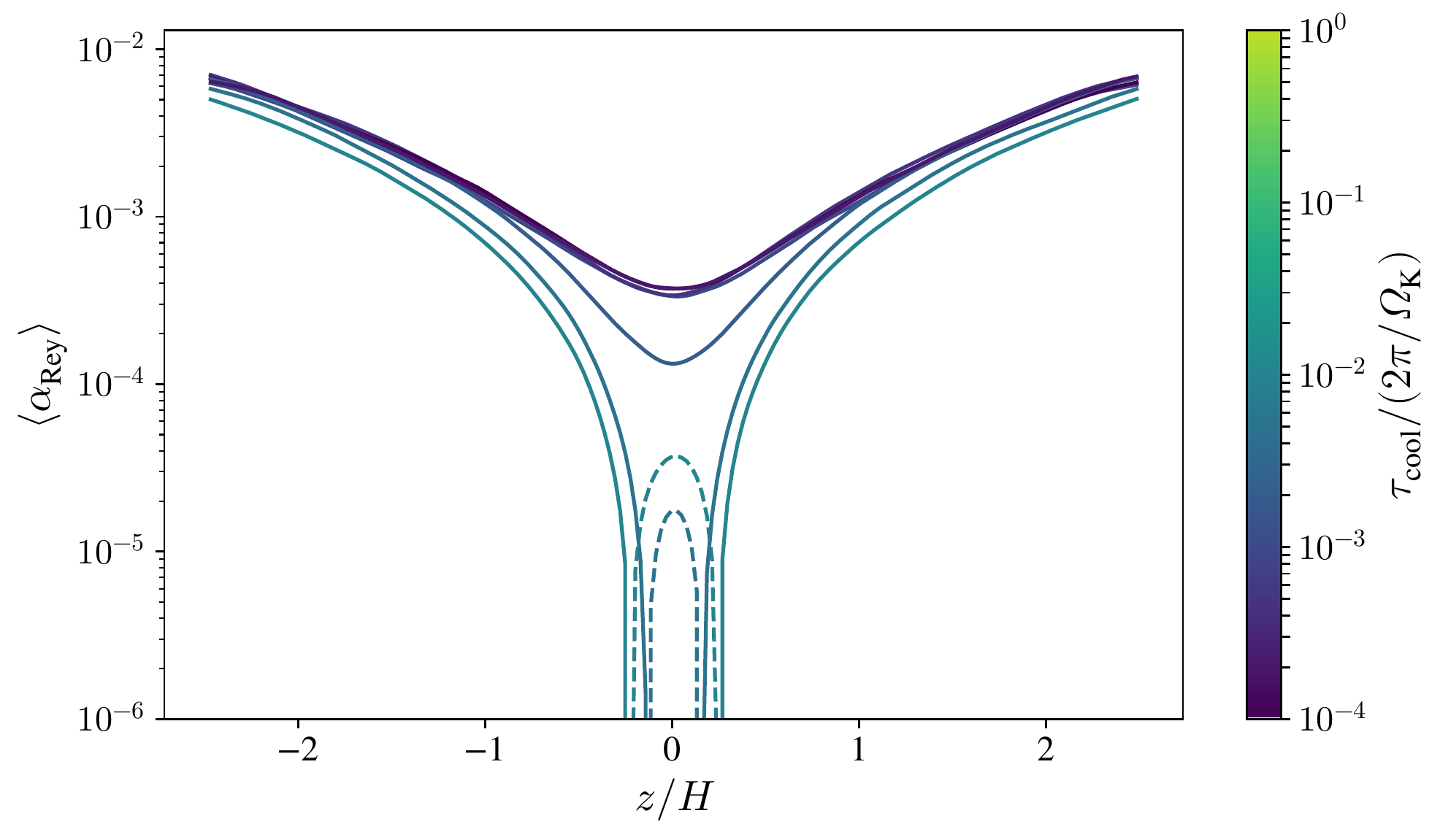}
    \caption{$\alpha$ as function of height for all models showing VSI growth. Averaging is performed in radial and azimuthal direction and between 200 and 500 orbits. Negative values are shown with dashed lines.}
    \label{fig:alphaZTau}
\end{figure}
A possible explanation can be found looking at figure \ref{fig:alphaZTau} where we show $\alpha$ as a function of height. There we see that for $\tcool \leq 10^{-3}$ orbits the vertical $\alpha$ profiles are similar for all simulations, but for larger values of  $\tcool$ $\alpha$ decreases more sharply towards the midplane, eventually turning negative. 
\review{We interpret this as the vertical extent unstable for VSI getting smaller with longer cooling times, with regions close to the midplane being first to hamper the VSI modes.} 

\review{
The reason is simple. The time scale criterion in \citet{Lin+Youdin2015} is defined for a height of H above the midplane. This means that layers below this height with lower values for the vertical shear (respectively vertical epicylclic frequency $\kappa^2_z$) can be stabilised by the stratification in entropy. Thus the VSI does not disappear instantaneously over all heights, but is pushed to higher layers above the midplane, which, with their lower gas densities, are only weakly contributing to the global stresses.}

\begin{figure}
    \centering
    \includegraphics[width=\columnwidth]{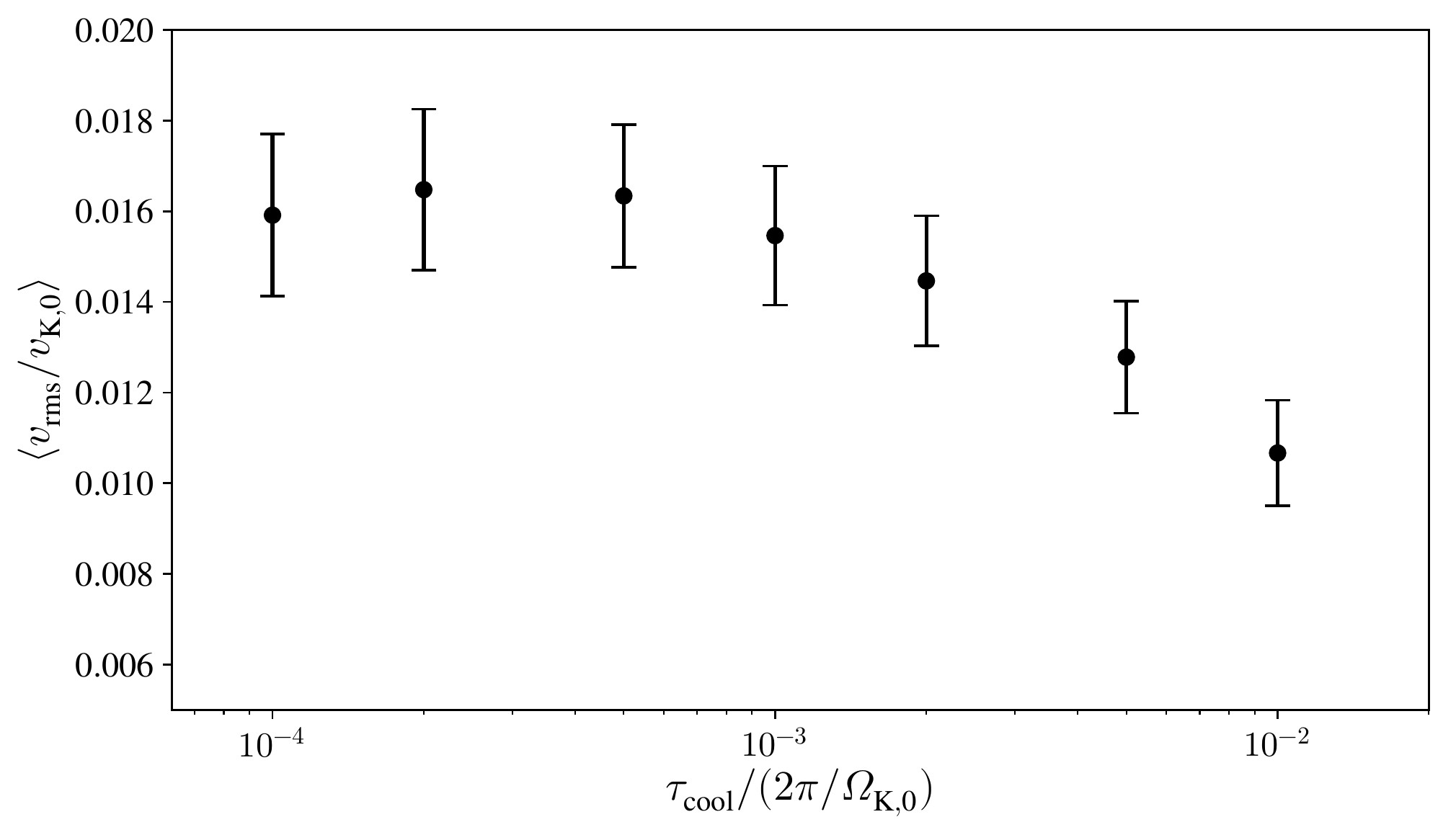}
    \caption{Space-time-averaged $v_\mathrm{rms}$ as function of $\tau_\mathrm{cool}$ for growing models. We observe a similar behaviour as in the lower plot of figure \ref{fig:alphaTau}.}
    \label{fig:vrmsTau}
\end{figure}

We also investigate the turbulent kinetic energy of the growing simulations by looking at the rms velocity:
\begin{equation}
v_\mathrm{rms}= \sqrt{\left(v_r -\langle v_r\rangle\right)^2 + \left(v_\theta -\langle v_\theta\rangle\right)^2+ \left(v_\varphi -\langle v_\varphi\rangle\right)^2} \: , \label{eqn:vrms}
\end{equation}
where brackets represent spatial averages in $\phi$ direction, and we assume that the averages $\langle v_r \rangle$ and $\langle v_\theta\rangle$ are equal to zero. We show the \review{mass-weighted averages over the simulation domain} for the simulations with $\tcool < \tau_\mathrm{cool,crit.}$ in figure \ref{fig:vrmsTserTau}. 
We find all simulations to grow to a saturated state within about 50 reference orbits. The initial growth rate for the simulations with $\tcool = 10^{-4}$ orbits is $\Gamma \approx 0.26 \Omega_K/2\pi$ and decreases to $\Gamma \approx 0.20 \Omega_K/2\pi$ for $\tcool = 10^{-2}$ orbits, which is lower but broadly consistent with the results obtained by \citet{Lin+Youdin2015}.

We also plot the total average of the rms velocity as a function of cooling time in figure \ref{fig:vrmsTau}. We find a similar decline in the rms velocity as observed for the $\alpha$ values in figure \ref{fig:alphaTau}.

\subsection{Dependence on temperature slope $q$}

\begin{figure}
    \centering
    \includegraphics[width=\columnwidth]{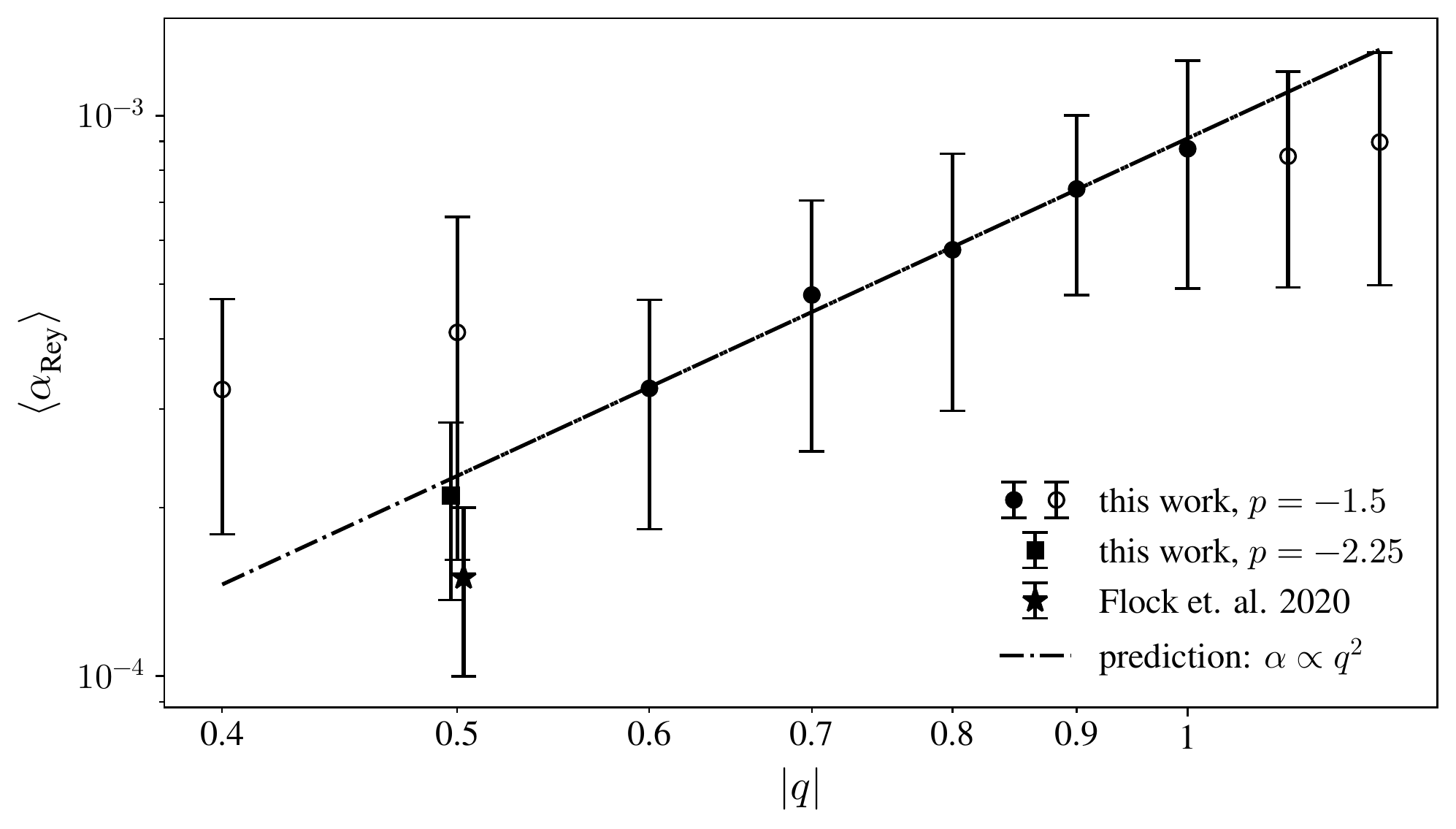}
    \caption{$\alpha$ averaged over the simulation space and between 200 and 500 orbits, where the error bars represent the standard deviation of the variations in time. The dot-dashed line shows our prediction from \citet{Manger+2020}. We indicate simulations that fit our prediction with filled and simulations that deviate significantly with empty circles. The square marker represents a simulation similar to q-0.5 but with a steeper midplane density gradient $p=-2.25 $ , and the starred marker represents the value from \citet{Flock+2020}.}
    \label{fig:alphaQ}
\end{figure}

Next we investigate the influence of the temperature slope $q$. We find that models with smaller \review{$|q|$} reach the saturated state later, but all simulations reach saturation within 150 orbits. We also observe a trend where larger \review{$|q|$} leads to larger saturated $\alpha$ values. Figure \ref{fig:alphaQ} shows this trend, where we plot the time average of $\alpha$ between 200 and 500 orbits. We find a 
clear correlation of $q$ and $\alpha$. We also plot our prediction $\alpha \propto q^2$ from \citet{Manger+2020}, but we find that this scaling is only compatible in the range \review{$0.6<|q|<1.0$}, as indicated by the filled markers. The empty markers represent simulations that deviate from our prediction and are discussed in more detail in \ref{sec:discussion}. We also show the value obtained by \cite{Flock+2020}, which is in agreement with our prediction when their larger azimuthal domain is considered.

\begin{figure}
    \centering
    \includegraphics[width=\columnwidth]{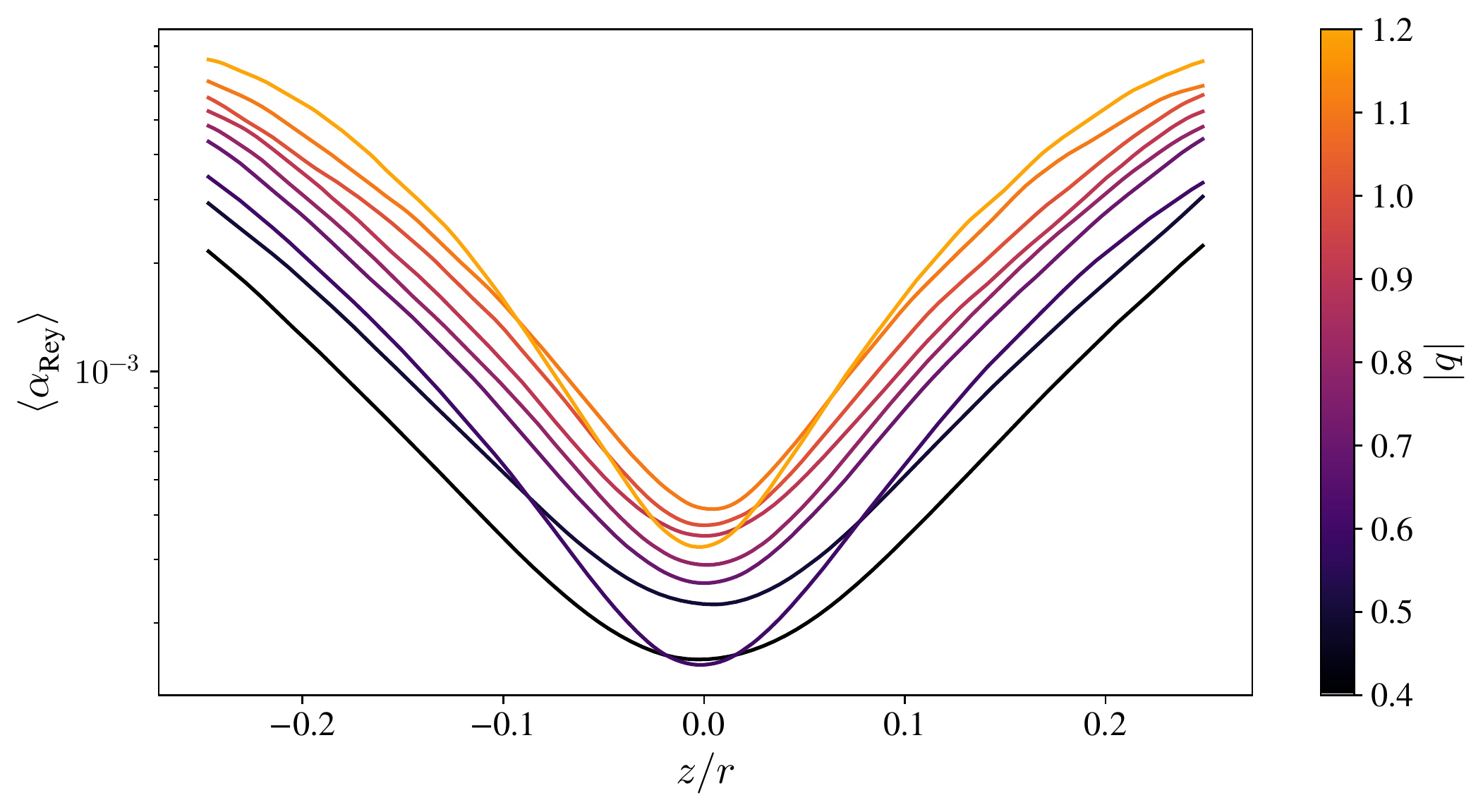}
    \caption{$\alpha$ as a function of height and averaged in radial and azimuthal direction and from 200 to 500 orbits. While $\alpha$ decreases monotonically as a function of $q$ away from the midplane, no such behaviour is observed near the midplane.}
    \label{fig:alphaZQ}
\end{figure}

 In figure \ref{fig:alphaZQ} we investigate $\alpha$ as a function of height $z$ above the midplane. We find that, while $\alpha$ decreases monotonically with \review{$|q|$} at heights larger than $|z| > 0.1 R_0$, there is no general scaling trend of $\alpha$ with $q$ near the midplane. This certainly influences the total values presented in figure \ref{fig:alphaQ} and explains in part the deviation from the expected scaling.

\begin{figure}
    \centering
    \includegraphics[width=\columnwidth]{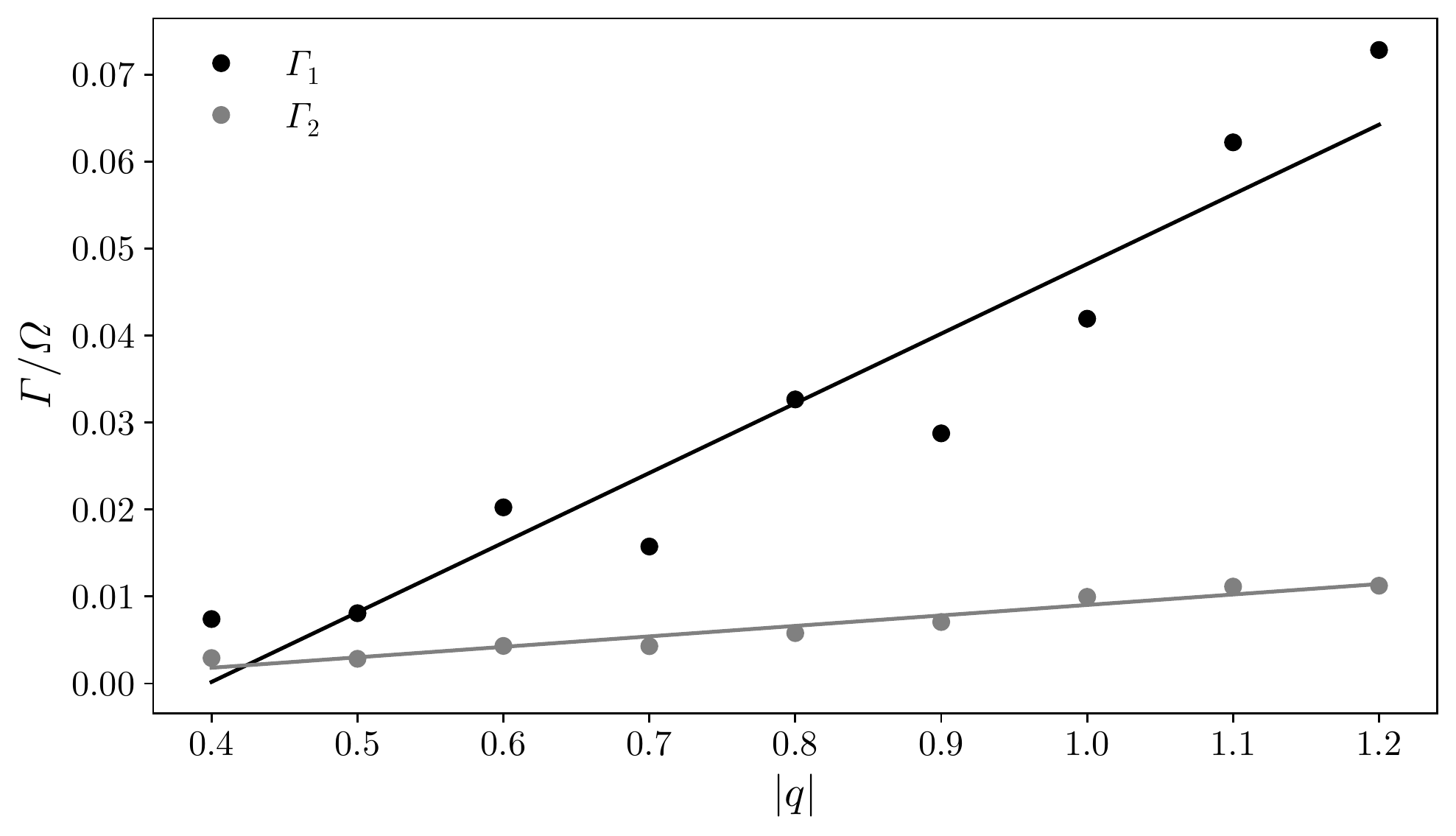}
    \caption{Initial (black) and secondary (grey) growth rates from figure \ref{fig:vrmsTserTau}. We also plot the linear relationship between $q$ and $\Gamma$ predicted by theory for both sets.}
    \label{fig:gammaQ}
\end{figure}
\begin{figure}
    \centering
    \includegraphics[width=\columnwidth]{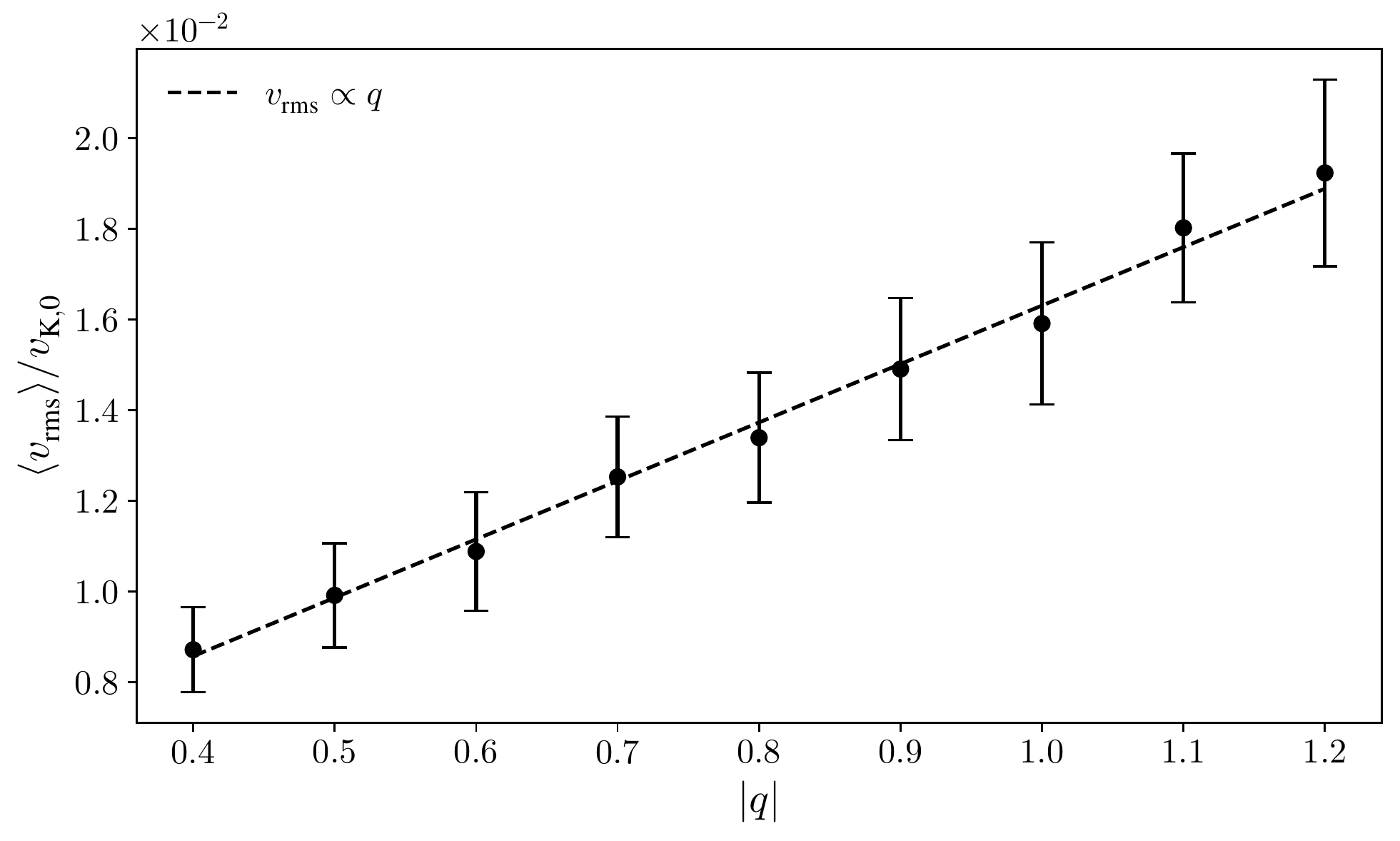}
    \caption{Rms velocity averaged over the simulation domain and between 200 and 500 orbits. A linear relationship between $v_\mathrm{rms}$ and $q$ is recovered.}
    \label{fig:vrmsQ}
\end{figure}

Next, we look again at the turbulent rms velocities of the simulations.
We find that the saturated $v_\mathrm{rms}$ has a linear dependence on $q$ in all simulations, as shown in figure \ref{fig:vrmsQ}. This result is in line with the expected scaling we derived in \citet{Manger+2020} (equation 14 therein), where we proposed a linear dependence of $v_\mathrm{rms}$ on the total vertical shear of the disk at 1 $ H $ and therefore a linear dependence on $q$. The evolution of $ v_\mathrm{rms}$  with time is provided in figure \ref{fig:vrmsTserQ}.

We also determined the growth rates for each simulations and show the results for both the primary and secondary growth phase in figure \ref{fig:gammaQ}. \review{In the primary phase, so-called finger modes start to grow from the upper parts of the disk, where the vertical shear is strongest. Once these modes reach the disk midplane, merging with their respective counterparts from the other side of the disk, the secondary growth phase of the now so-called body modes begins.} We additionally plot the theoretically predicted scaling \citep{Nelson+2013,Stoll+Kley2014}:
\review{
\begin{equation}
    \frac{\Gamma}{\Omega} \propto |q|\,\frac{H}{R}
\end{equation}
}
yielding a linear relationship between $\Gamma$ and $q$. This relationship is recovered for both the primary and secondary growth rates, though it is less evident in the former, as the short time spent in this regime gives only few data points to determine the growth rate from. 

\section{Discussion}
\label{sec:discussion}

\subsection{Dependence on cooling time $\tau_\mathrm{cool}$ }
Our investigation of the cooling time parameter revealed two distinct regimes within the parameter range allowing the VSI to grow. For $\tau_\mathrm{cool}\lesssim 10^{-3}$ local orbits, the cooling time does not significantly influence the strength of the turbulent angular momentum transport, which holds constant at $ \alpha \approx 10^{-3} $. At larger values, the total $ \alpha $ decreases with increasing cooling time to $ \alpha\approx 2\cdot 10^{-4}$ for $ \tau_\mathrm{cool} = 10^{-2} $, the longest cooling time supporting VSI growth in our simulations. This dichotomy is also observed in the average total kinetic energy. We therefore think that the decrease in $ \alpha $ occurs because the disk cannot extract as much energy at longer cooling times due to the stronger buoyancy forces generated in the disk. This leads to less turbulence being generated, leading to a lower $ \alpha $.
However, once the cooling time becomes sufficiently short, the disk becomes quasi-isothermal and a further decrease in cooling time does not influence the level of turbulence generated.  

\review{The assumption of a spatially uniform cooling time is a strong simplification, in reality cooling times vary with height above the mid-plane and distance to the star \citep[][]{Pfeil+Klahr2018}. Even the midplane of the disk might have a high optical depth and be stable for VSI modes, the upper atmosphere may provide sufficiently short cooling times to drive turbulence. Even further up the thermal decoupling of dust and gas at low densities stabilises the gas flow \citep[][]{Malygin+2017}. Thus ultimately only full radiation hydro models can provide definitive predictions on the occurrence of VSI in protoplanetary disks \citep[][]{Stoll+Kley2014}. But even there the latter effect of dust gas decoupling is neglected, and the only current remedy is using a 2D map for the cooling time similar to the ones used in \citet{Pfeil+Klahr2020}.}

Nevertheless, our general finding in this paper is consistent to \cite{Flock+2020}, which applied flux-limited diffusion for the transport of heat, who reported the lowest $ \alpha $ in the inner portion of their simulation $\alpha \approx 10^{-4}$, where due to higher optical depths the cooling times were longer than at larger radii with . They report a slight decrease in $ \alpha $ after a maximum is reached at around 30 AU $\alpha \approx 2 \cdot 10^{-4}$, while our results suggest no such decrease should occur. They also note that at larger radii the VSI might not have fully developed yet, possibly explaining the discrepancy.\review{ Our results are also consistent with the results for an irradiated disk presented in \citet[]{Stoll+Kley2014} and \citet[]{Stoll+Kley2016}, who reported $\alpha $ values in the range $ 0.5- 2 \cdot 10^{-4}$ and $ 1- 4 \cdot 10^{-4}$, though an exact comparison is difficult due to variations in both temperature slope and cooling time with radius. }

We therefore can safely expect the VSI to at least generate a viscosity gradient at the inner edge of its unstable zone, creating something akin to the MRI dead-zone edge described in non-ideal MHD simulations and opening further routes to triggering the RWI in a disk,\review{ as has been suggested by \citet{Flock+2020}}. Future studies should investigate the feedback loop between long cooling times in high surface density regions, which then automatically would be low viscosity zones, contrary to the general constant $\alpha$ models.

\subsection{Dependence on temperature gradient $q$ }
In figure \ref{fig:alphaQ} we show the total averaged $ \alpha $ value as a function of the temperature gradient $q$. We find that there is a correlation like the one reported by \citet{Pfeil+Klahr2020}, but only the values marked with black circles also follow the slope we predicted in our analysis in \citetalias{Manger+2020}. In contrast, the values marked with empty circles deviate from our prediction, the ones at $ q = 0.4 $ and $ 0.5 $ significantly. The same deviation is however not observed in the corresponding figure showing the rms velocity (figure \ref{fig:vrmsQ}), where the linear relationship between $ q $ and $v_\mathrm{rms}$ is recovered. Therefore, there has to be an additional effect of the local temperature profile on the turbulent angular momentum transport in those simulations. This is supported by the fact that the simulations with lower $q$ tend to have %falling instantaneous 
$\alpha$ values to decay over time after an initial saturation of turbulence is reached (figure \ref{fig:alphaTserQ}), though in some cases the values eventually rise again.

\begin{figure*}
    \centering
    \includegraphics[width=0.8\linewidth]{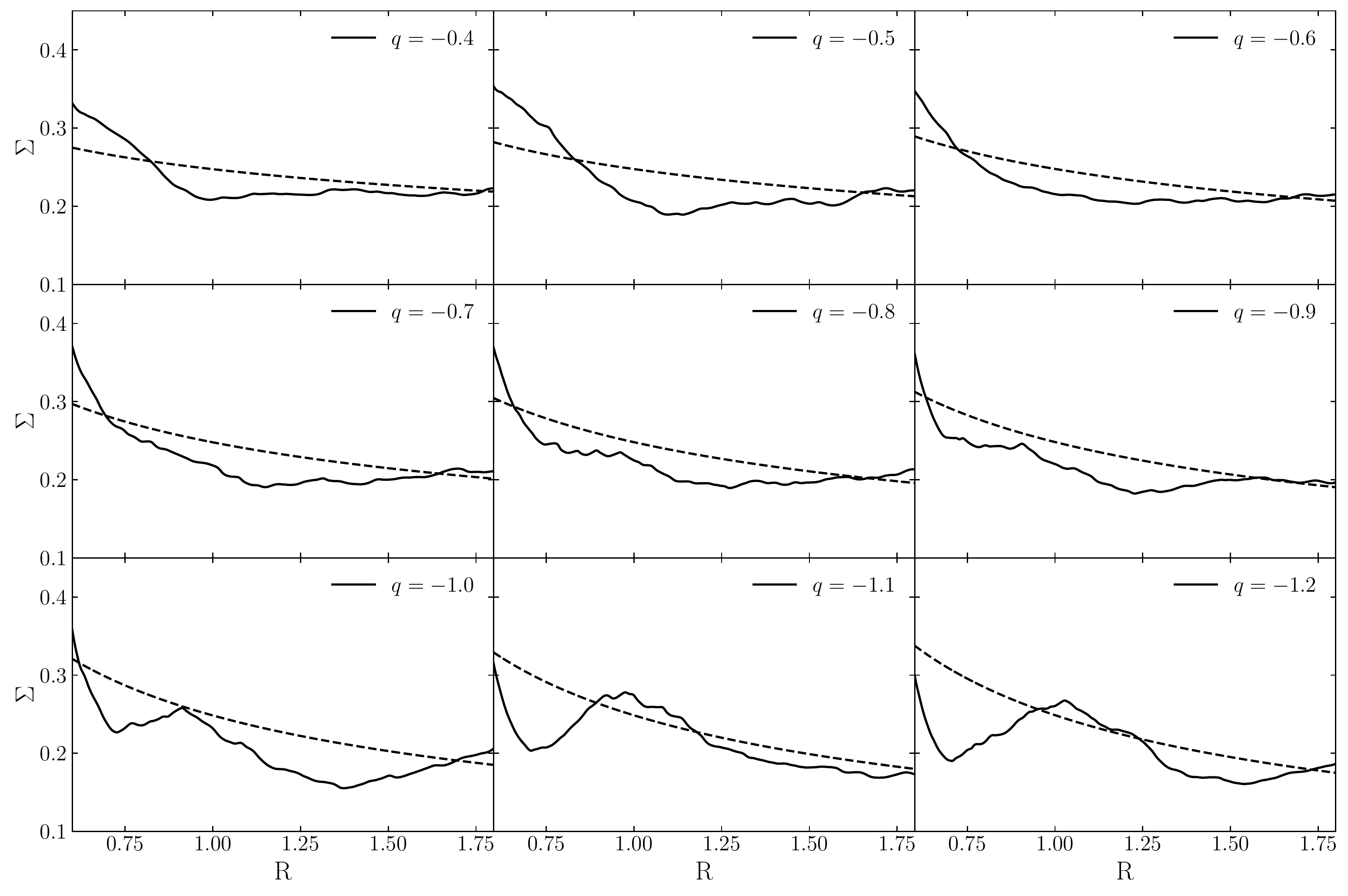}
    \caption{Initial (dashed) and final (solid line) column density $\Sigma$ of the gas for each simulation. We find that $\Sigma$ changes in all models, but we find a more extreme shift in $\Sigma$ for the models that do not fit our prediction in figure \ref{fig:alphaQ}.}
    \label{fig:SigmaQ}
\end{figure*}

The apparent enhanced turbulent viscosity can be explained if we consider that the disk adjusts to a new surface density profile. We plot the Column density  
\begin{equation}
    \Sigma (R) = \int_{-3H}^{3H} \rho \,  \mathrm{d}z
\end{equation}
for all simulations in figure \ref{fig:SigmaQ}.

While we find that $\Sigma$ changes to some degree in all our simulations due to mass accretion towards the central star 
, the change in surface density profile is stronger for simulations with $q=0.4$ and $0.5$, where $\alpha$ was higher than expected. In those cases, we observe a new, steeper surface density profile emerging during the simulation.

This reorganization can be explained by our choice to perform the parameter study with a fixed value $p=-1.5$ for the density gradient for all simulations, irrespective of the slope of the temperature gradient. However, the chosen value for the density gradient is strictly only compatible with the $q$ chosen in the fiducial case, as the disk demands a constant mass accretion rate to be maintained \review{\citep{Lynden-Bell+1974}}:
\begin{equation}
    \dot M \propto \Sigma \nu = \Sigma \alpha c_s H
\end{equation}
has to be constant throughout the disk. This leads to a requirement for the slope of the radial density gradient $p$:
\begin{equation}
    p = -3 -\frac{3}{2}q
\end{equation}

We performed an additional simulation for $q=-0.5$ with the density gradient $p=-2.25$ as required by the condition set by the mass accretion rate.
Marked with square in figure \ref{fig:alphaQ}, it shows that this disk now has a lower $\alpha$ that is in good agreement with the value we expect from the scaling law we proposed in \citetalias{Manger+2020} and we expect a similar outcome for the case with $q=0.4$.
This simulation result also compares well with the result presented in \citet[ star marker in figure \ref{fig:alphaQ}]{Flock+2020}, where a total average $\alpha$ of $1.4\cdot 10^{-4}$ is reported for a starlight-heated protoplanetary disk with comparable disk parameters. \review{Similar results have also been presented in \citet{BarrazaAlfaro+2021} for a simulation with an isothermal equation of state and $H/R =0.1$ and $q=-0.5$.}

\section{Conclusions}
In this work, we present the second part of our parameter study of the Vertical Shear Instability. Here we investigated the influence of the temperature slope $q$ and the thermal relaxation time $\tau_\mathrm{cool}$ on the evolution and non-linear saturated state of the VSI with special attention to the stresses generated within the disk. For all simulations we chose a density slope $p=-1.5$ and a disk aspect ratio of $H/R = 0.1$ \review{at the reference radius,} applicable to the outer regions of a protoplanetary disk.

We find that the VSI generates accretion stresses with stress-to-pressure ratios $\alpha$ in the range of $\alpha = 10^{-4} - 10^{-3}$, in line with the range of results presented in previous studies. We also find that $\alpha$ scales with the square of the slope $q$ of the temperature profile, confirming our theoretical prediction from \citetalias{Manger+2020}. With this we can also further explain the wide range in reported values for the stress-to pressure ratio $\alpha$ reported in previous works, which can be attributed to the differences in the inital setups chosen by the authors.

We also find that the VSI shows two distinct behaviours when changing the thermal relaxation time of the gas. For very short relaxation times, below $\tau_\mathrm{cool} \approx 10^{-3}$ orbits in this work, all simulations show $\alpha \approx 9\cdot 10^{-4}$ with no significant change with increasing $\tau_\mathrm{cool}$. Above this value however, $\alpha$ decreases with increasing $\tau_\mathrm{cool}$, and the longest relaxation time supporting the VSI ($\tau_\mathrm{cool} =  10^{-2}$ orbits in this work) shows an averaged $\alpha= 2\cdot 10^{-4}$. 

Such a general behavior of the VSI can naturally lead to dead-zone edges similar to the one found in models of the Magneto-rotational-instability and enable the formation of a vortex at the inner edge of the VSI active disk zone.
As an adaptive alpha model for the VSI we therefore propose:
\begin{equation}
    \alpha = 10^{-3} \left(\frac{h}{0.1}\right)^{2.6} q^2 S(\tau_{\mathrm{crit}} - \tau_{\mathrm{cool}})
\end{equation}
with $S$ the sigmoid function.

As thermal relaxation times vary with distance to the star and height above the midplane plus strongly depend on the local surface density and the effective opacity provided by the dust grains, the actual variation of $\alpha$ is an even more complex function involving further studies of full radiation hydrodynamics plus collisional thermal dust-gas coupling.

To accommodate a wider range of parameters, we chose to narrow the azimuthal extent of the disk model to $\Delta \phi = 45^\circ$. We are aware that this choice suppresses non-azimuthal global instabilities such as the Rossby-Wave-Instability as shown in \citet{Manger+Klahr2018}, and we will address the influence of the $q$ and $\tau_\mathrm{cool}$ on these effects in a future publication.
Furthermore, we also focused solely on the gas component of the protoplanetary disk, neglecting the dust present within these disks. Future studies should include the dust component and investigate the influence of the dust on the turbulent accretion stresses within the disk.

\section*{Acknowledgements}
The authors thank Wladimir Lyra for useful discussions.
This research was supported by the Deutsche Forschungsgemeinschaft Schwerpunktprogramm: SPP 1385 "The first ten million years of the Solar System" under contract KL 1469/4-(1-3) "Gravoturbulente Planetesimal Entstehung im fr\"uhen Sonnensystem", (SPP 1992 Exploring the diversity of extrasolar planets under contract KL 1469/17-1, KL 1469/16-1 and KL 1469/16-2, by SPP 1833 "Building a Habitable Earth" under contract KL 1469/13-1 \& KL 1469/13-2 "Der Ursprung des Baumaterials der Erde: Woher stammen die Planetesimale und die Pebbles? Numerische Modellierung der Akkretionsphase der Erde.", the Munich Institute for Astro- and Particle Physics (MIAPP) of the DFG cluster of excellence "Origin and Structure of the Universe and in part at KITP Santa Barbara by the National Science Foundation under Grant No. NSF PHY11-25915. 
T.P. and T.B. acknowledge the support of the German
Science Foundation (DFG) priority program SPP 1992
``Exploring the Diversity of Extrasolar Planets'' under grant
No. BI 1816/7-2.
The authors gratefully acknowledge the Gauss Centre for Supercomputing (GCS) for providing computing time for a GCS Large-Scale Project (additional time through the John von Neumann Institute for Computing (NIC)) on the GCS share of the supercomputer JUQUEEN \citep{Stephan:202326} at J\"ulich Supercomputing Centre (JSC). GCS is the alliance of the three national supercomputing centres HLRS (Universit\"at Stuttgart), JSC (Forschungszentrum J\"ulich), and LRZ (Bayerische Akademie der Wissenschaften), funded by the German Federal Ministry of Education and Research (BMBF) and the German State Ministries for Research of Baden-W\"urttemberg (MWK), Bayern (StMWFK) and Nordrhein-Westfalen (MIWF). Additional simulations were performed on the ISAAC clusters of the MPIA and the COBRA, HYDRA and DRACO clusters of the Max-Planck-Society, both hosted at the Max-Planck Computing and Data Facility in Garching (Germany). Further simulations were performed on the Rusty cluster of the Flatiron Institute. H.K. also acknowledges additional support from the DFG via the Heidelberg Cluster of Excellence STRUCTURES in the framework of Germany's Excellence Strategy (grant EXC-2181/1 - 390900948).

\section*{Data Availability Statement}
The data underlying this article were generated and accessed from large-scale computing clusters at the Max-Planck Computing and Data Facility and the Flatiron Institute and will be shared on reasonable request to the corresponding author.
%%%%%%%%%%%%%%%%%%%%%%%%%%%%%%%%%%%%%%%%%%%%%%%%%%


% The best way to enter references is to use BibTeX

\bibliographystyle{mnras}
\bibliography{references} % if your bibtex file is called example.bib

%\newpage

\appendix
\section{Resolution study}
To validate our results, we performed simulations at twice the resolution in each direction for three of the cases considered, namely the fiducial run tau1e-4 and the runs tau1e-2 and q-0.5. 

We find our fiducial resolution in good agreement with the results obtained at higher resolution, and are confident that our chosen resolution is sufficient to capture the VSI. The later but faster growth found in the high resolution cases can be explained by the more stable initial density profile used in these runs (see e.g. equation 12 in \cite{Nelson+2013}) and the additional modes accessible due to the higher resolution employed.

When examining the $\alpha$ values however we find that the case q-0.5 shows a lower $\alpha$ value in the high resolution simulation. This indicates that the higher resolution case restructures the disk differently than the fiducial resolution cases discussed in the main text. It also supports our finding that the $\alpha$ values found for low $ q $ values in figure \ref{fig:alphaQ} are too high and that our prediction is still valid.

\begin{figure}
    \centering
    \includegraphics[width=\columnwidth]{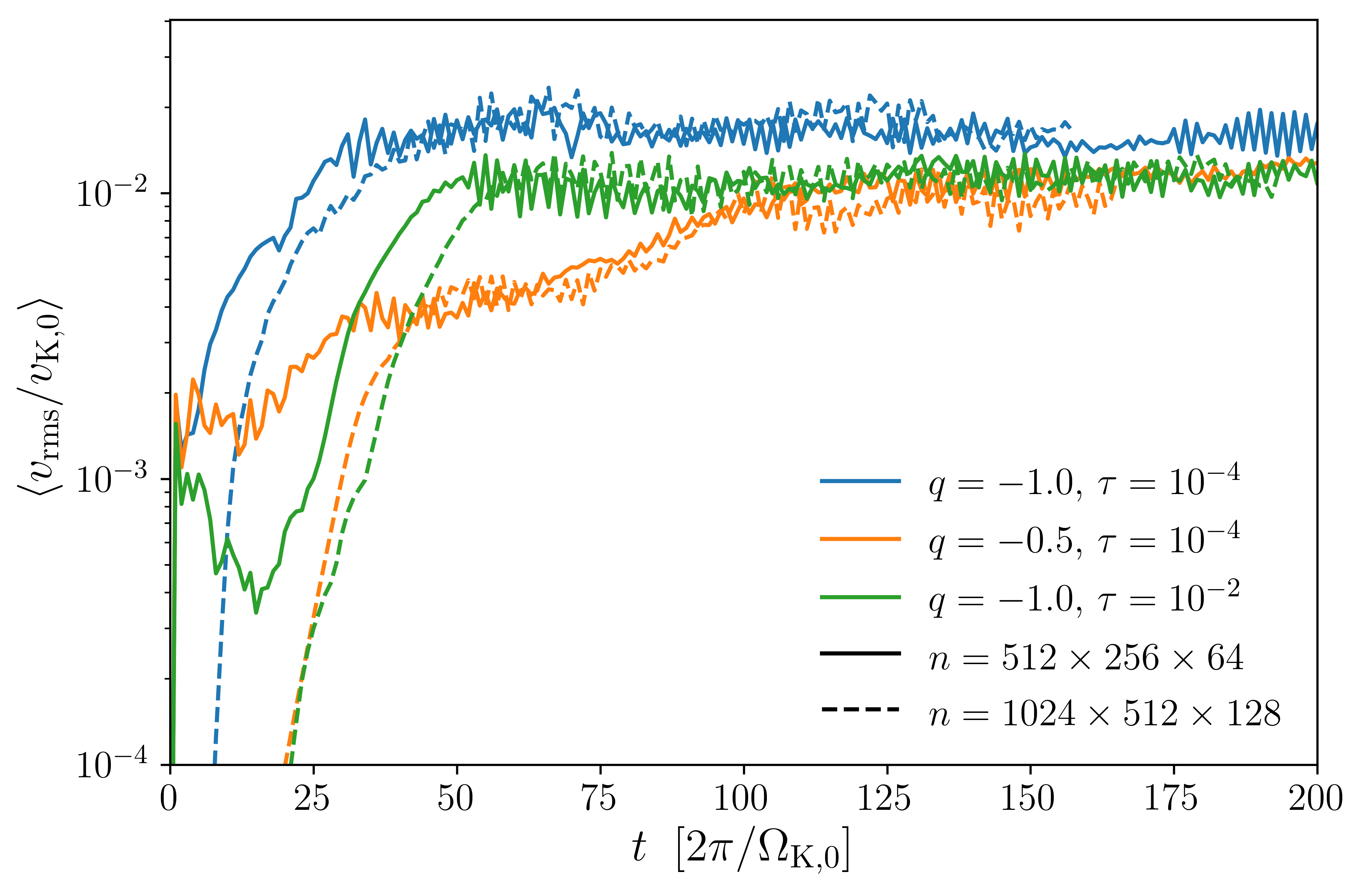}
    \caption{Rms velocity as a function of time for the fiducial and double resolution cases. We find good agreement between both resolutions for all cases during the saturated phase.}
    \label{fig:res_vrms}
\end{figure}

\begin{figure}
    \centering
    \includegraphics[width=\columnwidth]{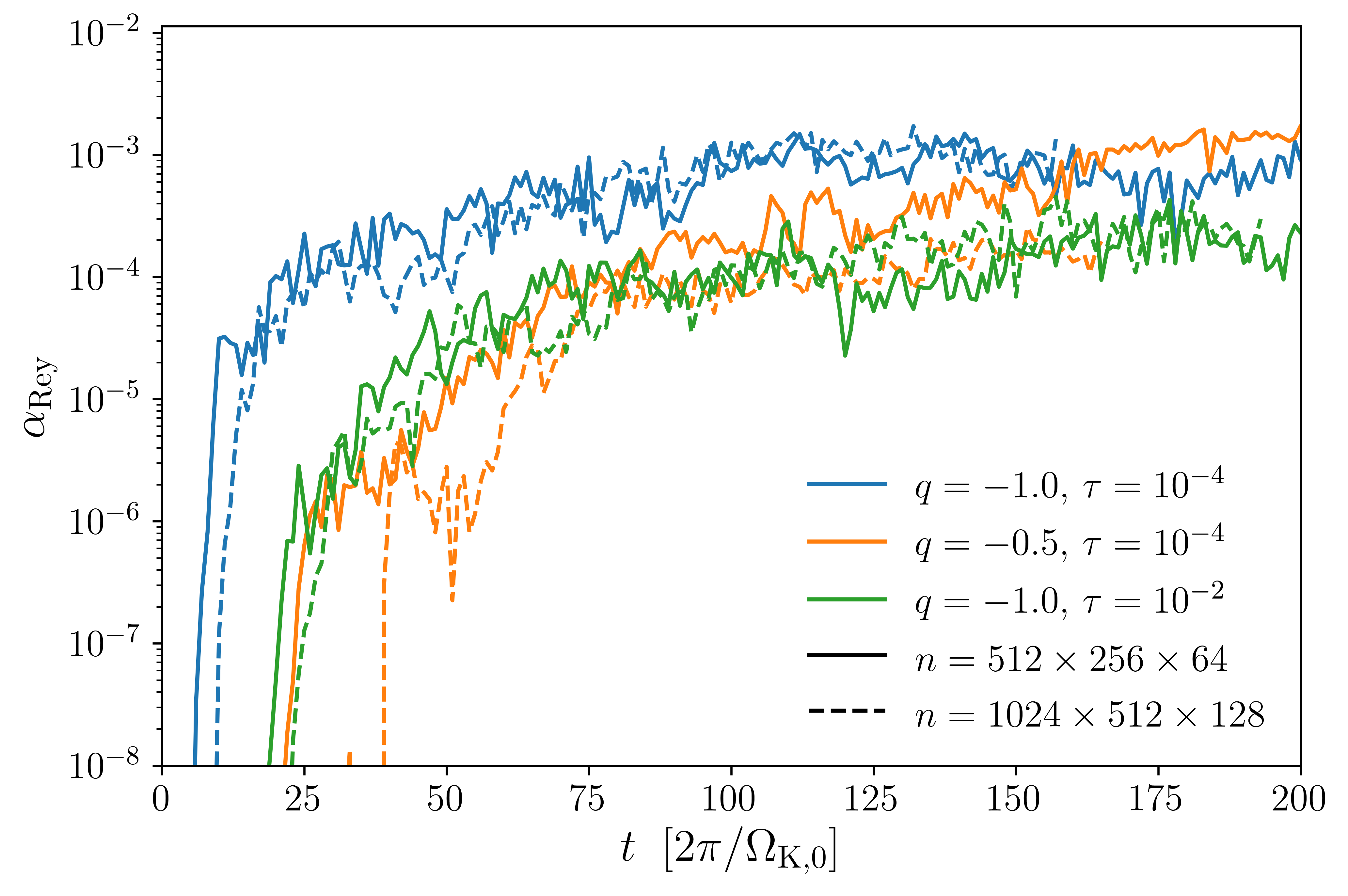}
    \caption{Reynolds-stress parameter $\alpha$ as a function of time for the fiducial and double resolution. Good agreement is found for the simulations with $ q = -1.0$, but we find a significantly lower $\alpha$ for the higher resolution case for $ q= -0.5$.}
    \label{fig:res_alpha}
\end{figure}

\section{Additional figures}
\begin{figure}
    \centering
    \includegraphics[width=\columnwidth]{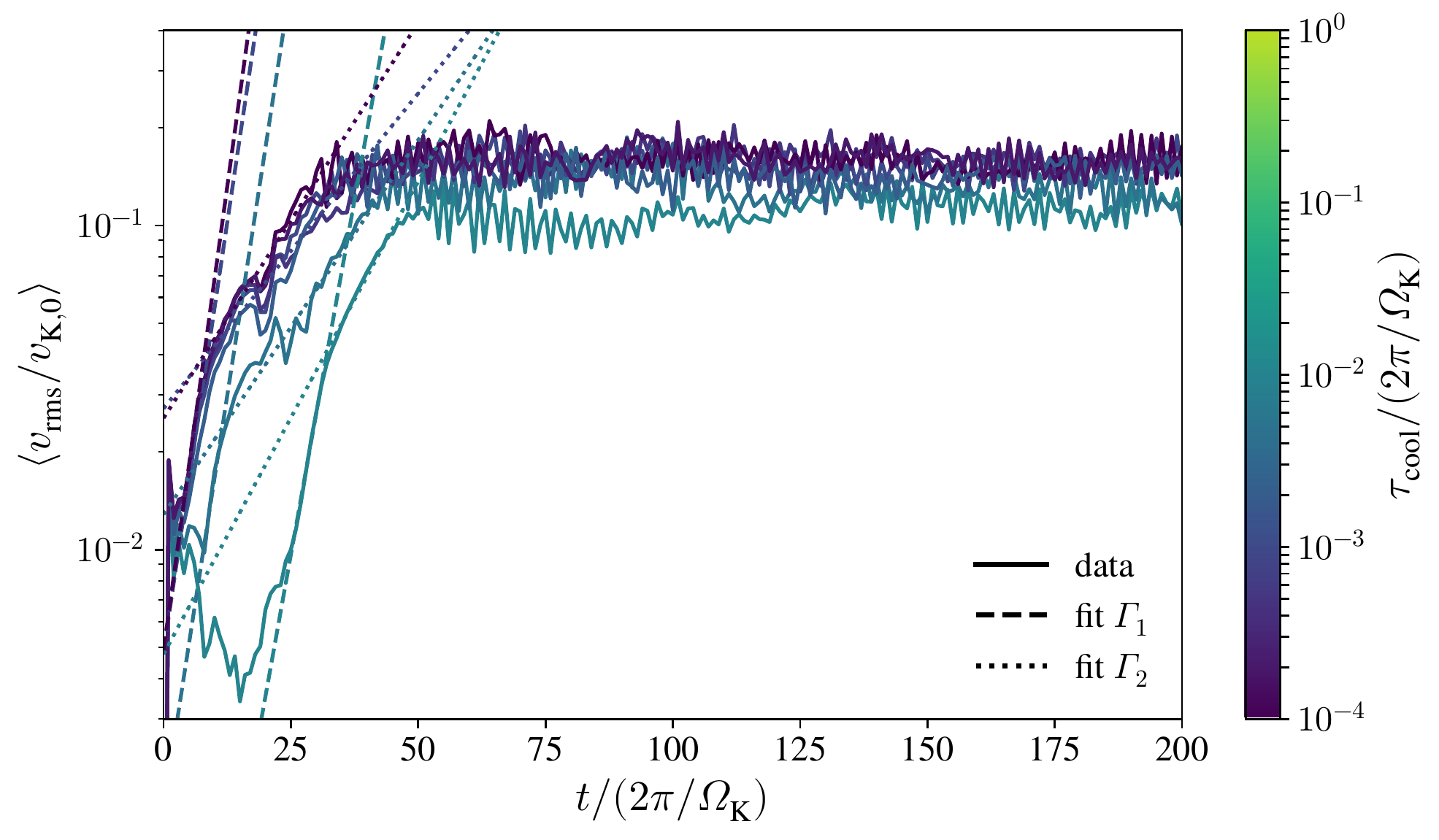}
    \caption{Space averaged $v_\mathrm{rms}$ as function of time for all simulations with $\tau_\mathrm{cool} \leq 10^{-2}$. The dashed lines show a fit to the initial growth phase, the dotted lines show the secondary growth phase. }
    \label{fig:vrmsTserTau}
\end{figure}

\begin{figure}
    \centering
    \includegraphics[width=\columnwidth]{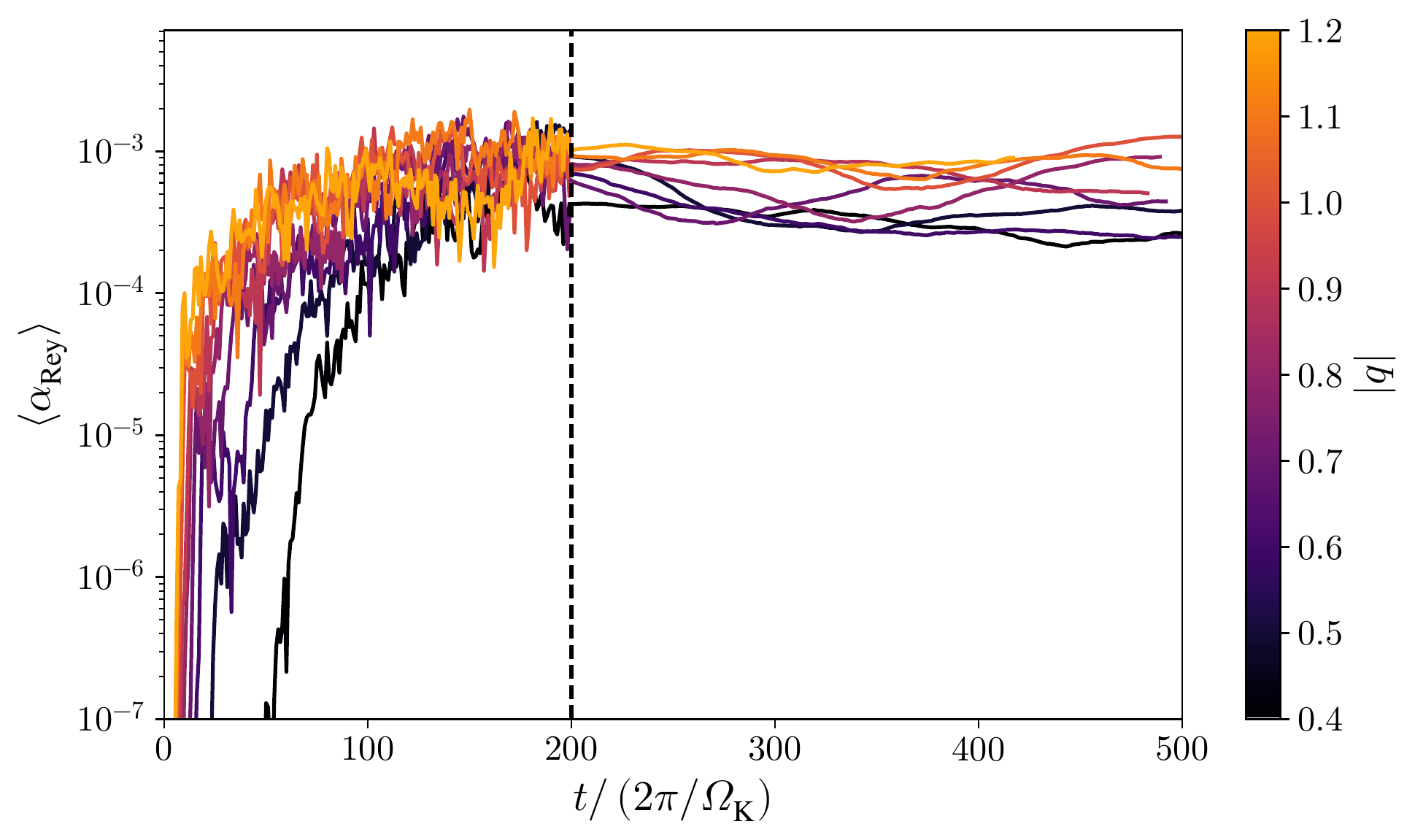}
    \caption{Spatially averaged $\alpha$ as a function of time. After 200 orbits (dashed line) we switch to a moving average with a width of 100 orbits in time to highlight the overall trend of $\alpha$ during the saturated phase. The color represents the temperature slope $q$ of the respective simulation.}
    \label{fig:alphaTserQ}
\end{figure}

\begin{figure}
    \centering
    \includegraphics[width=\columnwidth]{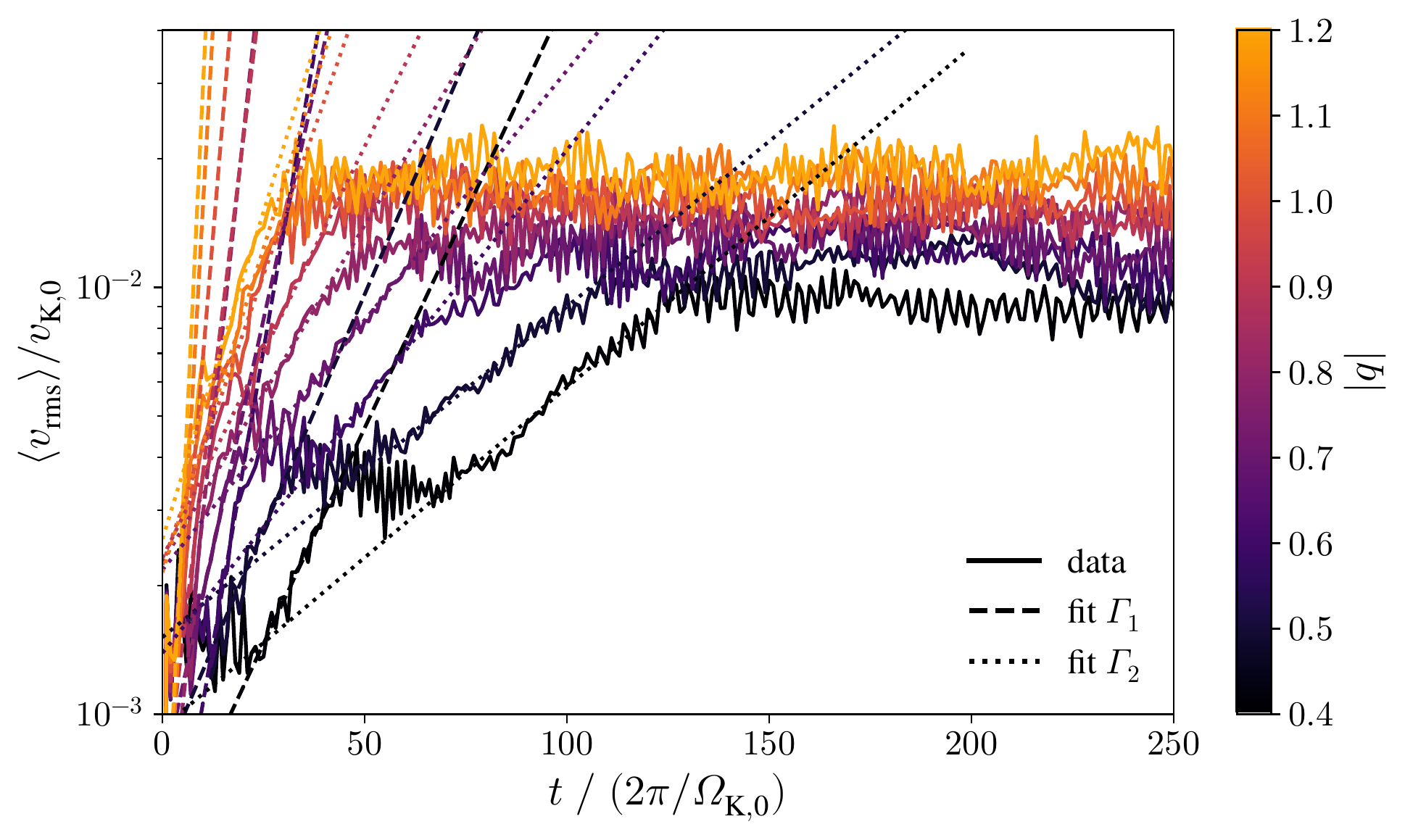}
    \caption{Averaged rms velocity as a function of time for different values of q. The dashed line shows the fit to the initial growth period, while the dotted line shows the fits to the secondary growth phase. The growth rates are shown in figure \ref{fig:gammaQ}.}
    \label{fig:vrmsTserQ}
\end{figure}

% Don't change these lines
\bsp	% typesetting comment
\label{lastpage}
\end{document}